\documentclass[conference]{IEEEtran}
\IEEEoverridecommandlockouts

\usepackage{cite}
\usepackage{amsmath,amssymb,amsfonts}
\usepackage{algorithmic}
\usepackage{graphicx}
\usepackage{textcomp}
\usepackage{xcolor}
\usepackage{soul} 
\usepackage{threeparttable} 

\def\BibTeX{{\rm B\kern-.05em{\sc i\kern-.025em b}\kern-.08em
    T\kern-.1667em\lower.7ex\hbox{E}\kern-.125emX}}

\usepackage{fancyhdr}
\usepackage{url}
\fancypagestyle{ewfirst}{%
  \fancyhf{}

  \fancyfoot[C]{\parbox{\textwidth}{%
    \centering\fontsize{7}{8}\selectfont
    Authors' accepted version. Published by VDE VERLAG in the proceedings of European Wireless 2025, pp. 91--96.\\
    Published version: \url{https://ieeexplore.ieee.org/document/11442741}.
  }}
}

\begin{document}

\title{Interoperability Assessment of Fronthaul and E2 Interfaces for Multi-Vendor Open RAN Systems\\
\thanks{This work was supported by the Department for Science, Innovation and Technology (DSIT).}
}

\author{\IEEEauthorblockN{Farhad Mehran, Charles Turyagyenda, and Dritan Kaleshi}
\IEEEauthorblockA{\textit{Digital Catapult}\\
London, UK\\
\{farhad.mehran,\,charles.turyagyenda,\,dritan.kaleshi\}@digicatapult.org.uk}
}

\maketitle
\thispagestyle{ewfirst}

\begin{abstract}
Open Radio Access Networks (RANs) aim to transform the existing wireless network architectures by adopting open interfaces, disaggregating and virtualising RAN components, and enabling dynamic control of the system via data-driven methods. To accelerate the development of interoperable RAN solutions, rigorous assessment of Open Fronthaul and E2 interfaces based on O-RAN Alliance Interoperability Testing (IOT) specifications plays a key role as the integration of these interfaces seems to be very challenging, e.g. due to the need for alignment of Fronthaul IOT profiles and unforeseen challenges for E2 interface integration as Near-Real-Time RAN Intelligent Controller (Near-RT RIC) is relatively a new component introduced within the RAN architecture. In this paper, selected parts of the latest findings and lessons learned from interoperability assessment of Open Fronthaul and E2 interfaces for 13 multi-vendor Open RAN systems deployed in SmartRAN Open Network Interoperability Centre (SONIC) Labs are presented.\end{abstract}


\section{Introduction}
\begin{figure}[t]
\centerline{\includegraphics[scale=0.33]{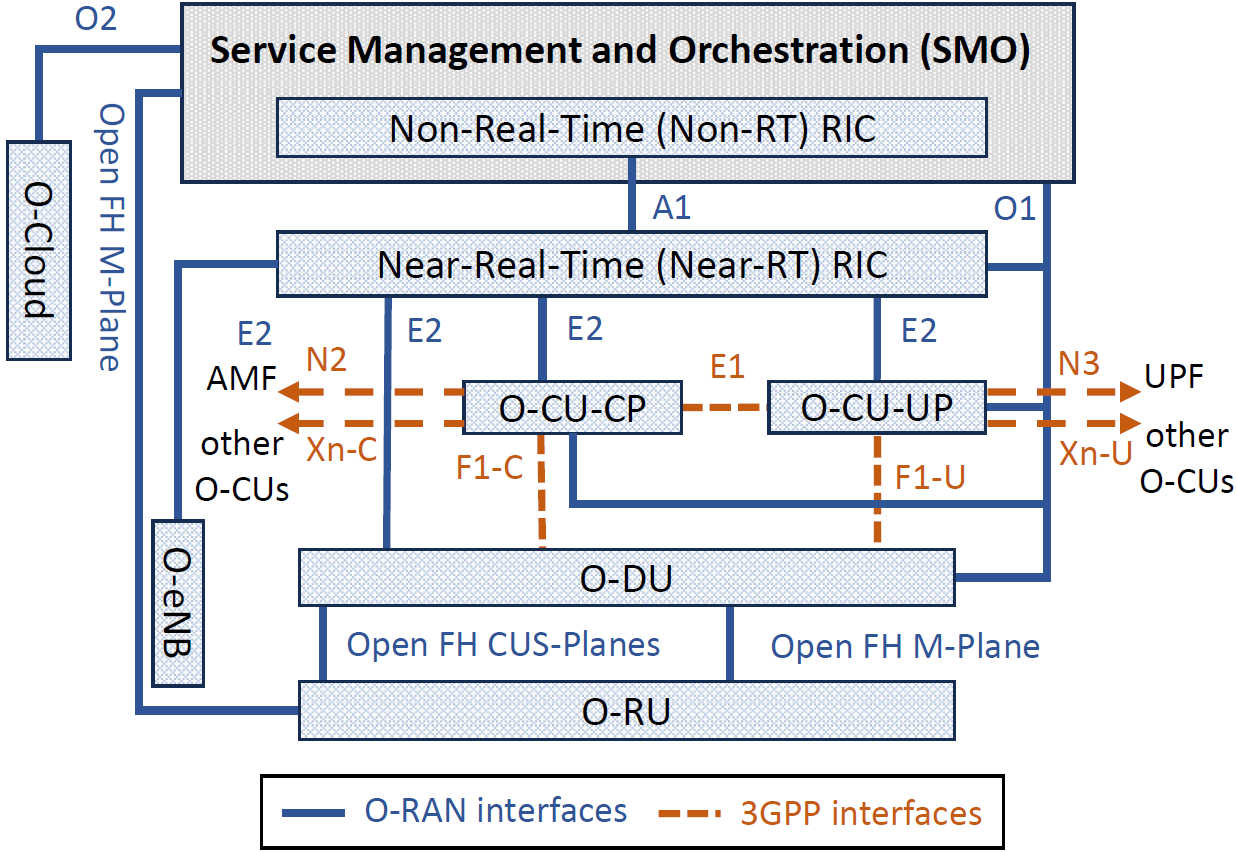}}\caption{O-RAN architecture \cite{oranarchitecture}: RAN
components are interconnected via O-RAN interfaces (solid lines) and 3GPP
interfaces (dashed lines).}
\label{openrandiagram}
\end{figure}

Open Radio Access Network (RAN) has emerged as an architectural overhaul for the future wireless cellular networks to create a competitive ecosystem with a diverse range of RAN products \cite{concise}. This transformative evolution allows to mix-and-match innovative solutions, prevents vendor lock-in, and enables data-driven intelligent RAN operation. Still, Open RAN paradigm inevitably impose new challenges that did not exist in traditional monolithic systems. More specifically, one of the key architectural principles of Open RAN is openness for the interfaces that are interconnecting the constituent components of the disaggregated RAN. O-RAN Alliance defines the specifications for the Open RAN architecture, interfaces, and interoperability test cases between the pairs of RAN components. Fig. \ref{openrandiagram} shows the interfaces for Open RAN as defined by its embodiment in O-RAN specifications \cite{oranarchitecture}. This architecture embraces the 3GPP functional split 7.2 \cite{3gppsplit}, and disaggregates the RAN functions into O-RAN Central Unit (O-CU), Distributed Unit (O-DU), and Radio Unit (O-RU). 

Among the O-RAN interfaces, integration and interoperability validation of Open Fronthaul (FH) interface attracted strong interest due to the fact that the O-RU not only accounts for significant portion of CAPEX/OPEX for Mobile Network Operators (MNOs) but it also plays a key role in determining the overall performance of the network, e.g. by changing the MIMO configurations to scale the throughput performance or improve the coverage at the cell edge. Integration of Open Fronthaul interface is a challenging task, particularly for multi-vendor cases, due the need for alignment of several configuration parameters within the IOT profiles of the O-RU and O-DU, e.g. for delay management, IQ compression, etc.

Also, within the O-RAN architecture, Near-Real-Time RAN Intelligent Controller (Near-RT RIC) is a core component as it empowers network optimisations and in this regard, E2 interface is the key interface that connects the O-DU/O-CU to the Near-RT RIC. The data-driven applications hosted within Near-RT RIC receive Key Performance Measurements (KPMs) from the O-DU/O-CU over the E2 interface, use them in pre-trained models, and send back control actions to the O-DU/O-CU. Hence, to ensure the correct operation of the Near-RT RIC (and functionality of its applications), it is essential to assess the interoperability of the E2 Nodes with respect to O-RAN standard specifications. However, for enabling interoperability over multi-vendor E2 interfaces, the readiness of the E2 interfaces for the existing O-RAN products has not yet reached the state of maturity. This is one of the major bottlenecks that currently does not allow to unleash the full potential of RIC-enabled use cases. This is due to several factors, namely lack of maturity in O-RAN standard specifications, proprietary elements within different vendors' implementations, and lack of clarity about E2 design requirements as Near-RT RIC is relatively new component within RAN domain.

To assess the reality of Open RAN technologies, validate maturity of products, and accelerate the developments needed for enabling plug-and-play solutions, various integration and testing centres around the globe serve as commercially neutral testbeds for conducting experimental evaluations and providing impartial assessment for the level of openness and interoperability that the Open RAN products have. In this regard, up to now, SmartRAN Open Network Interoperability Centre (SONIC Labs) led the collaborative integration and interoperability testing for 47 Open RAN systems (created from products of 21 vendors) on its commercially neutral multi-tenancy open network testbed \cite{SONICLabs}. 

Although Open RAN ecosystem has made significant progress towards defining test specifications and methodologies for performance validations, there is a notable gap in available literature about the technical findings, lessons learned, and analysis of observations as a result of conducting interoperability assessments between pairs of Open RAN products. To fill this gap, in \cite{coh1to4}, we presented the key findings from the interoperability and E2E performance evaluations of 34 Open RAN systems deployed in SONIC Labs, e.g. lessons learned on interoperability issues for KPM verification and reporting on E2 interface, alignment of FlexRAN implementations to successfully integrate Open Fronthaul interface, etc. In this paper, we share parts of the new findings and lessons learned as a result of conducting interoperability assessment of 13 Open RAN systems that were recently integrated in SONIC Labs and were not covered in our previous work \cite{coh1to4}. In particular, we will, 
\begin{itemize}
  \item Provide the results and breakdown of test cases that were selected and designed according to O-RAN Alliance specifications for assessing interoperability of Open Fronthaul and E2 interfaces,
  \item Present guidelines for designing delay management profiles for O-RU and O-DU to enable successful integration and interoperability over Open Fronthaul interface, and 
  \item Share insights on E2 interface validations, e.g. verification of nested Information Elements (IEs) within IOT testing procedures and its impact on the overall interoperability between the pairs of RAN components.   
\end{itemize}

\section{System Setup In SONIC Labs}
\begin{figure}[t]
\centerline{\includegraphics[scale=0.4]{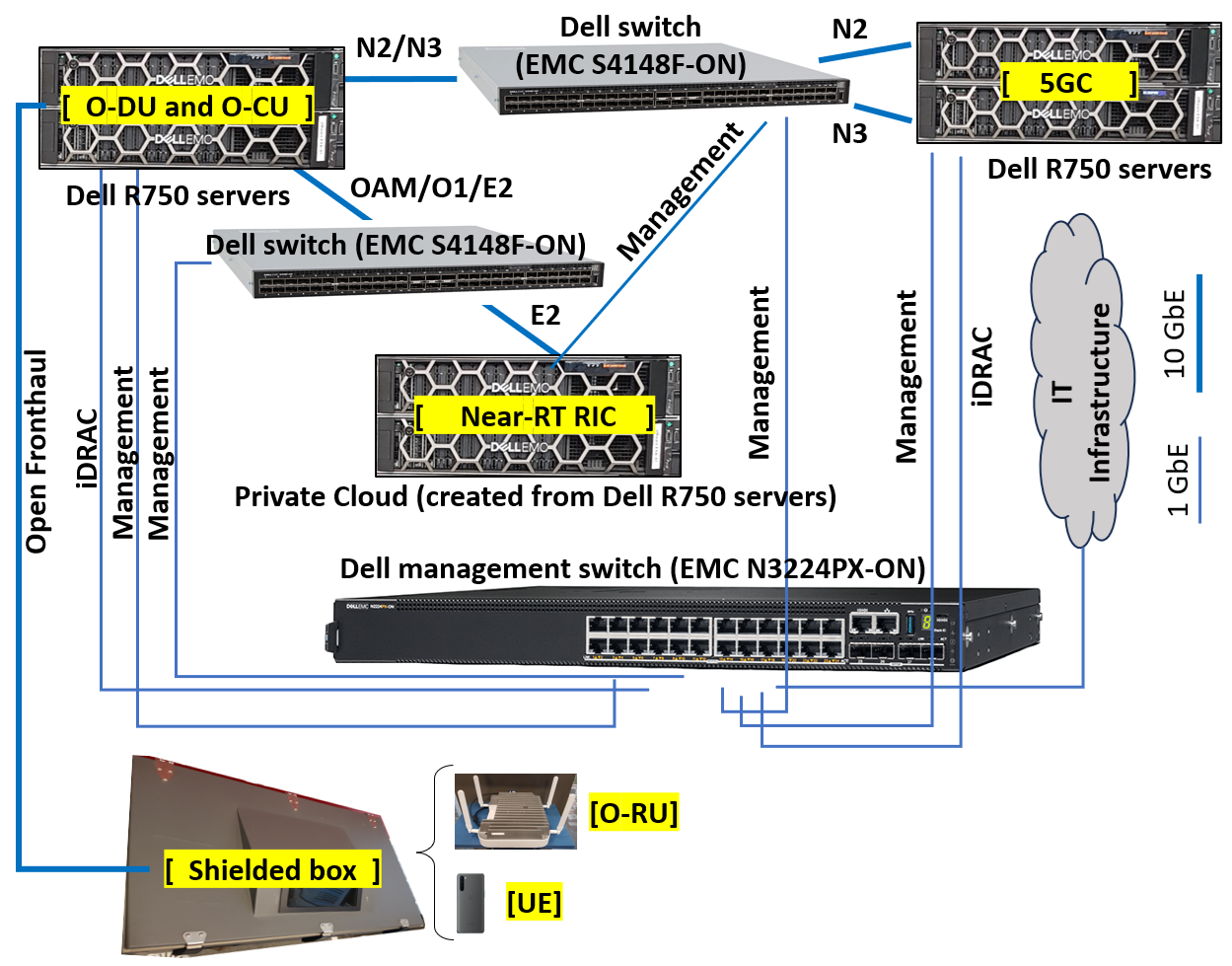}}\caption{High-level architecture for an instance of Open RAN system deployed in SONIC Labs.}
\label{sonicinfra}
\end{figure}

This section describes the high-level architecture and the test setups for the Open RAN systems deployed on SONIC Labs testbed. Fig. \ref{sonicinfra}  illustrates the hardware infrastructure for an instance of Open RAN system, which was composed of Dell R750 servers for deploying O-DU, O-CU, Near-RT RIC, and 5G Core (5GC) solutions, Dell switches to interconnect the Open RAN components, and also shielded box that hosts the O-RU and the commercial UE. For all of the systems, the O-RU was operating in 5G\,NR\,n77 band with 100\,MHz bandwidth and configured as 4x4 MIMO.
For the Open Fronthaul interoperability testing, the System-Under-Test (SUT) was composed of O-DU and O-RU. For the interoperability assessment of E2 interface, the SUT was composed of O-DU and Near-RT RIC, and also O-CU and Near RT-RIC.

\begin{table*}[h!]
\caption{Open Fronthaul Interoperability Validations for two open RAN systems.}
\label{signaling}
\begin{tabular}
{|p{33pt}|p{44pt}|p{90pt}|p{168pt}|p{48pt}|p{48pt}|}
\hline
\hline
\textbf{Test Case}&	\textbf{Functionality}	&\textbf{O-RAN Test Specification} &\textbf{Test Case Description} &  \textbf{Open RAN \#1}& \textbf{Open RAN \#2}\\
\hline
\hline
$2.2.2.1$&
S-Plane&O-RAN.WG4.IOT.0-R004-v12.00\,\cite{wg4iot}&Functional test of O-DU + O-RU using ITU-T G.8275.1 profile (LLS-C1)&N/A&Pass\\
\hline
$2.2.2.3$&
S-Plane&O-RAN.WG4.IOT.0-R004-v12.00\,\cite{wg4iot}&Functional test of O-DU + bridged network + O-RU using ITU-T G.8275.1 profile (LLS-C3)&Pass&N/A\\
\hline
$2.2.2.8$&
S-Plane&O-RAN.WG4.IOT.0-R004-v12.00\,\cite{wg4iot}&Performance test of O-DU + Two O-RUs using ITU-T G.8275.1 profile (LLS-C1)&N/A&Pass\\
\hline
$2.2.2.10$&
S-Plane&O-RAN.WG4.IOT.0-R004-v12.00\,\cite{wg4iot}&Performance test of O-DU + bridged network + Two O-RUs using ITU-T G.8275.1 profile (LLS-C3)&Pass&N/A\\
\hline
$2.2.3.1$&
CU-Planes&O-RAN.WG4.IOT.0-R004-v12.00\,\cite{wg4iot}&Radio Layer 3 C-Plane establishment and Initial Radio U-Plane data transfer&Pass&Pass\\
\hline
$2.2.3.2$&
CU-Planes&O-RAN.WG4.IOT.0-R004-v12.00\,\cite{wg4iot}&Radio U-Plane downlink data transfer (Downlink throughput performance)&Pass&Pass\\
\hline
$2.2.3.3$&
CU-Planes&O-RAN.WG4.IOT.0-R004-v12.00\,\cite{wg4iot}&Radio U-Plane uplink data transfer (Uplink throughput performance)&Pass&Pass\\
\hline
$2.2.4.5$&
CU-Planes&O-RAN.WG4.IOT.0-R004-v12.00\,\cite{wg4iot}&Delay Management \#3, normal fronthaul latency&Pass&Pass\\
\hline
\hline
\end{tabular}
\label{FHIOT}
\end{table*}

\section{Interoperability Assessment: Challenges and Lessons Learned}

\subsection{Open Fronthaul Interface}
Table \ref{FHIOT}
presents a summary of the results of interoperability assessment for the Open Fronthaul interface of two Open RAN systems. The set of test cases was designed according to the O-RAN IOT test specifications within Work Group 4 (WG4) \cite{wg4iot}. This includes the assessment of interoperability and functionality of Open Fronthaul interface for Control and User Planes (CU-Planes) and Synchronisation Plane (S-Plane). Validation of Management Plane (M-Plane) was not part of the system integration for these systems.  
 
\textit{Interoperability Assessment for CU-Planes:} the O-RAN functional split 7.2x has stringent delay requirements for Fronthaul interface. To integrate the CU-Planes, it is essential to ensure that the transmission and reception windows at the O-DU are aligned with the delay characteristics of transport network and O-RU. For this alignment, the delay management model as defined in O-RAN WG4 Control, User and Synchronisation Plane (CUS) specification \cite{wg4cus} is used. 

\begin{figure}[t]
\centerline{\includegraphics[scale=0.52]{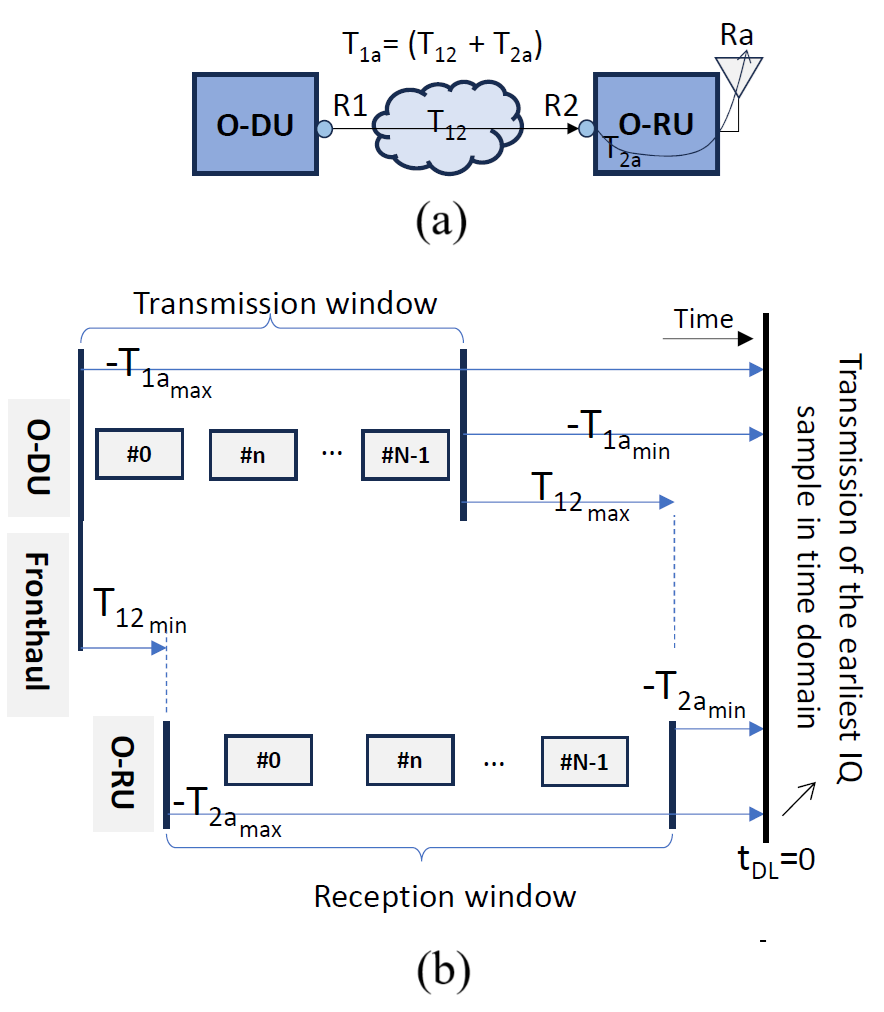}}\caption{Delay management model for the O-RU and O-DU for DL transmissions: (a) reference points, (b) timing relations (taken from \cite{wg4cus,ecpri}).}
\label{timingdelay}
\end{figure}

The focus of this paper is on DL transmissions, and the related delay model that defines the reference points for input and output of the equipment for Open Fronthaul interface and the related timing relations are shown in Fig. \ref{timingdelay}. 
Here, $T_{2a}$, $T_{1a}$, and $T_{12}$ define the delay from receive interface at O-RU ($R2$) to antenna interface at O-RU ($Ra$), delay from transmit interface at O-DU ($R1$) to $Ra$, and DL transport delay, respectively. The reference point for delay management ($t_{DL}=0$) is when the earliest IQ sample is transmitted from antenna interface and therefore, the transmission and reception windows are defined relative to $Ra$ and relations are worked out backward in timeline. For the CU-Planes transmissions, the transmission window at the O-DU is defined by $T_{1a_{max}}$ and $T_{1a_{min}}$ that denote the earliest and latest transmission from the O-DU, respectively, and has the size of $W_t=(T_{1a_{max}}-T_{1a_{min}})$. Also, the reception window is specified by the $T_{2a_{max}}$ and $T_{2a_{min}}$ that denote the earliest and latest reception at the O-RU, respectively, which has the size of $W_r=(T_{2a_{max}}-T_{2a_{min})}$. Hence, the delay relationships for U-Plane DL transmission are given by \cite{wg4cus}
\begin{equation}
\left\{ \begin{aligned} 
T_{1a_{max},up}\leq (T_{2a_{max},up} + T_{12\,min})\\
T_{1a_{min},up}\geq (T_{2a_{min},up} + T_{12\,max})
\end{aligned} \right.
\label{eq1}
\end{equation}
\noindent and for the C-Plane, it follows 
\begin{equation}
\left\{ \begin{aligned} 
T_{1a_{max},cp}\leq (T_{2a_{max},cp} + T_{12\,min})\\
T_{1a_{min},cp}\geq (T_{2a_{min},cp} + T_{12\,max})
\end{aligned} \right.
\label{eq2}
\end{equation}
where subscripts $up$ and $cp$ present U-Plane and C-Plane timing parameters, respectively. From (\ref{eq1}) and (\ref{eq2}), the transmission/reception windows for U-Plane and C-Plane are denoted by $W_{t,up}/W_{r,up}$ and $W_{t,cp}/W_{r,cp}$, respectively. 

\begin{figure}[t!]
\centerline{\includegraphics[scale=0.44]{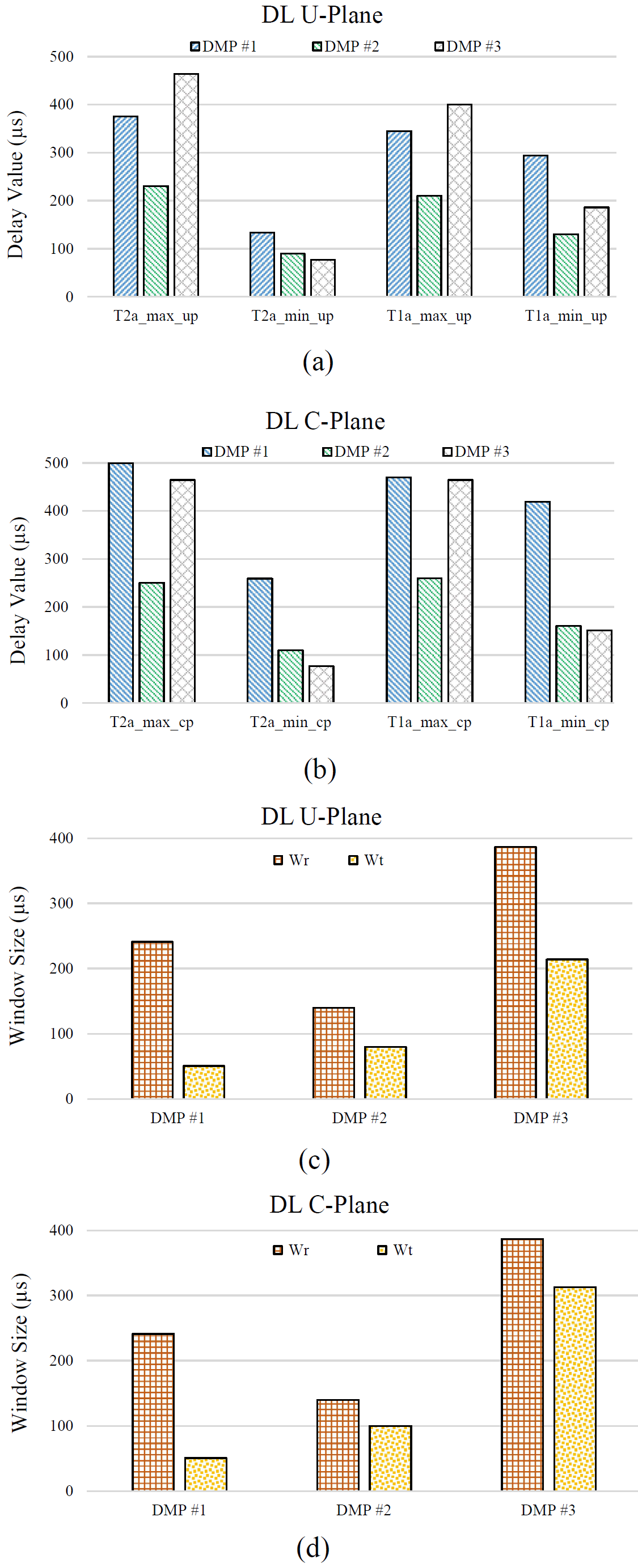}}\caption{Delay Management Profiles (DMPs) on Open Fronthaul for DL transmissions: Timing parameters for (a) U-Plane and (b) C-Plane; and size of transmission and reception windows for (c) U-Plane and (d) C-Plane. }
\label{fhdelay}
\end{figure}

To attain the information required for the O-DU to align the transmission window with the O-RU's reception window and transport network delay variations, O-RAN specifications \cite{wg4cus} introduced methods which are primarily relying on pre-defining and measuring the timing parameters. For this work, the O-DU timing parameters were configured based on pre-defined values (according to O-RU delay profile information) and without relying on Open FH M-Plane. Fig. \ref{fhdelay} presents examples of Delay Management Profiles (DMPs) configured and used within distinct Open Fronthaul interfaces that were successfully integrated for Open RAN systems. From this figure, the following observations can be made:\\ 
\indent\textit{(i) Design of Transmission $(W_t)$ and Reception $(W_r)$ Window Size:} for all the three cases, $W_r>W_t$ holds for both C-Plane and U-Plane, as inequalities (1) and (2) imply that the reception window size is configured to be equal or larger than the transmission window size and the potential variations of delay in transport network. Here, for DMP\,\#1, having $W_{r,up}>4\cdot W_{t,up}$ and $W_{r,cp}>4\cdot W_{t,cp}\,$(i.e. $W_{r,up},W_{r,cp}=241\,\mu s$ and $W_{t,up},W_{t,cp}=51\,\mu s$) provides high flexibility for where within the specified window the O-DU may commence the transmission, while limiting the reception window to $2\cdot W_{t,up}>W_{r,up}$ as in the DMP\,\#2 (with $W_{r,up}=140\,\mu s$ and $W_{t,up}=80\,\mu s$) provides smaller margins and flexibility.\\  
\indent\textit{(ii) Flexibility for Configuring Individual Delay Parameters:} for the DL transmission, the O-DU shall transmit the messages early and late enough to ensure the messages are received at the O-RU within the designed $W_r$, e.g. for DMP\,\#1, the O-DU shall transmit U-Plane messages before $T_{1a_{min},up}=294\,\mu s$ to ensure the message is received at the O-RU before $T_{2a_{min},up}=134\,\mu\,s$. Also, while the packets are transmitted late enough at the O-DU (i.e. after $T_{1a_{max},up}=345\,\mu s$) to ensure they are received within the designed $W_{r,up}$, $T_{2a_{max},up}$ can be configured in a way to relax the starting point of reception window of the O-RU and enlarge the $W_{r,up}$, e.g. for all the three systems, the O-DU sets $T_{2a_{max},up}>\,T_{1a_{max},up}$ to allow flexibility for processing early packets at the O-RU, and also helps with smooth interoperability testing, but it comes at the expense of addition processing cycles and increased buffer size that need to be optimised for efficient operation of network in real deployments. While increasing $T_{2a_{max},up}$ allows to enlarge $W_{r,up}$ and resulting in flexibility for early packets, the size of $W_{r,up}$ can not be further increased via relaxing $T_{2a_{min},up}$ for late packets, as O-RU inherently requires some time for generation of time domain IQ samples that starts from the reception of packets at the O-RU input point $R2$ and the transmission of the IQ samples from the antenna at $Ra$.

\indent\textit{(iii) Coordination of Timing Between C-Plane and U-Plane:\,}for processing U-Plane messages, the corresponding C-Plane messages shall be available. Thus, O-RAN specifications \cite{wg4cus} define the timing advance, $T_{cp{\textunderscore} adv\textunderscore dl}$, which specifies the amount of time that C-Plane messages shall arrive at the O-RU in advance of the latest possible time the first corresponding U-Plane messages may arrive, e.g. for DMP\,\#1, the timing advance was set to $T_{cp{\textunderscore} adv\textunderscore dl}=125\,\mu\,s$, with $T_{1a_{max},cp}=470\,\mu\,s$ and
$T_{1a_{max},up}=345\,\mu s$ configured, hence $T_{1a_{max},cp}\geq 
T_{1a_{max},up}+
T_{cp{\textunderscore}adv\textunderscore dl}$ holds and as a result, this configuration allowed sufficiently early transmission of C-Plane packets at the O-DU. Also, as $T_{2a_{max},cp}=500\,\mu s$ and
$T_{2a_{max},up}=375\,\mu\,s$ were set, $T_{2a_{max},cp}\geq 
T_{2a_{max},up}+
T_{cp{\textunderscore}adv\textunderscore dl}$ holds and therefore, the defined values allowed sufficiently early reception of C-Plane packets for the corresponding U-Plane packets at the O-RU. 

Also, for managing the delay between the C-Plane and U-Plane, it can be observed that for all the three systems, $W_{t,cp}\geq W_{t,up}$ holds with $T_{1a_{max},cp}>T_{1a_{max},up}$, allowing transmission of C-Plane packets earlier than the U-Plane packets. Still, not all systems were fully aligned with delay management model requirements specified in O-RAN CUS-Planes specifications \cite{wg4cus}, as in the case of DMP\,\#3, the implementation was based on giving priority to sending packets as soon as possible at the O-DU, hence focusing more on tuning $T_{1a_{max},cp}$ and $T_{1a_{max},up}$ rather than $T_{1a_{min},cp}$ and $T_{1a_{min},up}$. Hence, with $T_{cp{\textunderscore} adv\textunderscore dl}=0\,\mu s$, the integration and interoperability assessment for this implementation was successful over ideal Fronthaul transport network.\\  
\indent\textit{(iv) Deployment Requirements:\,}considering the design of transmission and reception windows, while DMP\,\#2 allowed the maximum transport delay to be $T_{12\,max}=40\,\mu s$, DMP\,\#1 accommodated a more flexible transport delay variations within $[0,160]\,\mu s$ that supports deployments that require less stringent requirements for delay variations of Fronthaul transport network.

\textit{Interoperability over S-Plane:} O-RAN WG4\cite{wg4cus} defines the topologies for supporting O-RU synchronisation needs, including the Lower-Layer Split (LLS) Configuration 1 (LLS-C1) in which O-DU directly synchronises the O-RU, and LLS-C3 that O-RU is synchronised via the Fronthaul network that distributes the network timing towards the O-RU and O-DU. The selection of synchronisation topology depends on various design factors, and is primarily specified by the deployment needs. For the deployment of Open RAN systems at SONIC Labs, as shown in Table \ref{FHIOT}, LLS-C1 and LLS-C3 were used and when one of these configurations were used for one system, the other configuration was not deployed and therefore the related IOT test cases were not applicable.

\begin{table*}[h!]
\caption{E2 Interoperability Validations (for both O-DU and O-CU) for the two open RAN systems.}
\label{signaling}
\begin{tabular}
{|p{32pt}|p{50pt}|p{90pt}|p{177pt}|p{47pt}|p{47pt}|}
\hline
\hline
\textbf{Test\,Case}&	\textbf{Functionality}	&\textbf{O-RAN\,Test\,Specification} &\textbf{Test Case\,Description} &  \textbf{Open\,RAN\,\#1}& \textbf{Open\,RAN\,\#2}\\
\hline
\hline
$7.2.2.1.1$&E2 on O-DU&O-RAN.WG3.E2TS-R003-v02.00\,\cite{wg3iot}&For the O-DU, RIC Subscription and RIC Indication procedures with REPORT Service Style\,1, Single Action Format Type\,1, Indication Message Format Type\,1&N/A&Pass\\
\hline
$7.2.2.1.1$&E2 on O-CU&O-RAN.WG3.E2TS-R003-v02.00\,\cite{wg3iot}&For the O-CU, RIC Subscription and RIC Indication procedures with REPORT Service Style\,1, Single Action Format Type\,1, Indication Message Format Type\,1&N/A&Pass\\
\hline
$7.2.2.5.1$&E2 on O-DU&O-RAN.WG3.E2TS-R003-v02.00\,\cite{wg3iot}&For the O-DU, RIC Subscription Delete procedure with RIC Event Trigger Style 1&N/A&Pass\\
\hline
$7.2.2.5.1$&E2 on O-CU&O-RAN.WG3.E2TS-R003-v02.00\,\cite{wg3iot}&For the O-CU, RIC Subscription Delete procedure with RIC Event Trigger Style 1&N/A&Pass\\
\hline
$7.24$&
E2 on O-DU
&O-RAN.WG8.IOT.0-R004-v12.00
\,\cite{wg8iot}&Verify O-DU successfully establish E2 Interface with near RT-RIC&Pass&Pass\\
\hline
\text{$7.24^{*}$}&
E2 on O-CU&O-RAN.WG8.IOT.0-R004-v12.00\,\cite{wg8iot}&Verify O-CU successfully establish E2 Interface with near RT-RIC&Pass&Pass\\
\hline
$7.25$&E2 on O-DU&O-RAN.WG8.IOT.0-R004-v12.00\,\cite{wg8iot}&Verify near RT-RIC server sends subscription request with type as REPORT to O-DU&Pass&Pass\\
\hline
\text{$7.25^{*}$}&E2 on O-CU&O-RAN.WG8.IOT.0-R004-v12.00\,\cite{wg8iot}&Verify near RT-RIC server sends subscription request with type as REPORT to O-CU&Pass&Pass\\
\hline
$7.27$&E2 on O-DU&O-RAN.WG8.IOT.0-R004-v12.00\,\cite{wg8iot}&Verify O-DU sends periodic report to the subscribed features to near RT-RIC server&Pass&Pass\\
\hline
\text{$7.27^{*}$}&E2 on O-CU&O-RAN.WG8.IOT.0-R004-v12.00\,\cite{wg8iot}&Verify O-CU sends periodic report to the subscribed features to near RT-RIC server&Pass&Pass\\
\hline
\hline
\end{tabular}
\label{E2IOT}
\begin{tablenotes}
       \item [1] *\,Developed based on this test case that was originally developed for O-DU within O-RAN specifications \cite{wg8iot}.
\end{tablenotes}
\end{table*}

\subsection{E2 Interface}

\textit{Interoperability Verifications for E2 Interface:} Table \ref{E2IOT} presents a summary of the results of IOT assessment for two systems. The test cases cover validations of two E2 Application Protocol (E2AP) modules:\,(\textit{i}) Global Procedures, e.g. E2 Setup procedure that verifies the exchange of basic configuration parameters between the E2 Nodes and Near-RT RIC, and (\textit{ii}) RIC Functional Procedures, such as RIC Subscription procedure, RIC Delete procedure, and RIC Indication procedure (e.g. to report KPM(s) per subscription request).      

\textit{Backward Compatibility of E2SM-KPM Specifications:} 
for the integration of E2 interface, the verifications began with assessment of Global Procedures. Initially, the \textit{E2 Setup Request} sent from the E2 Nodes were not decodable at the Near-RT RIC as the implementation of the O-DU and O-CU for E2SM-KPM were based on E2SM-KPM v1 specification \cite{e2smkpmv1}, while the implementation of Near-RT RIC from another vendor was based on E2SM-KPM v2.02 \cite{E2SMKPMv202}. It was found that these specifications for the E2SM-KPM were not backward compatible and therefore, the E2 Setup procedure was only successfully established after new E2 Agent instances were provided for implementation of O-DU and O-CU network functions to support E2SM-KPM v2.03\cite{E2SMKPMv203}, while the Near-RT RIC implementation of E2SM-KPM v2.02 was upgraded based on a new KPI monitoring (KPIMON) application.  

\textit{Alignment of Network Function Implementations According to O-RAN Specifications:} the assessment of successful E2 integration and proper interoperability via this interface is not limited to verification of individual signalling procedures between the E2 Nodes and Near-RT RIC, and it also requires to examine whether the messages are decoded correctly or not to ensure the defined functions are operating as expected. In some cases, these verifications are not straightforward due the complexity of data relationships within Information Elements (IEs), such as complex structure of nested IEs which are IEs that consist of other IEs inside them, e.g. following the establishment of communication over E2 interface in previous example, it was observed that the E2 Node was reporting incorrect 5G NR Cell Identity (NCI) to the Near-RT RIC. 

\begin{figure}[t]
\centerline{\includegraphics[scale=0.58]{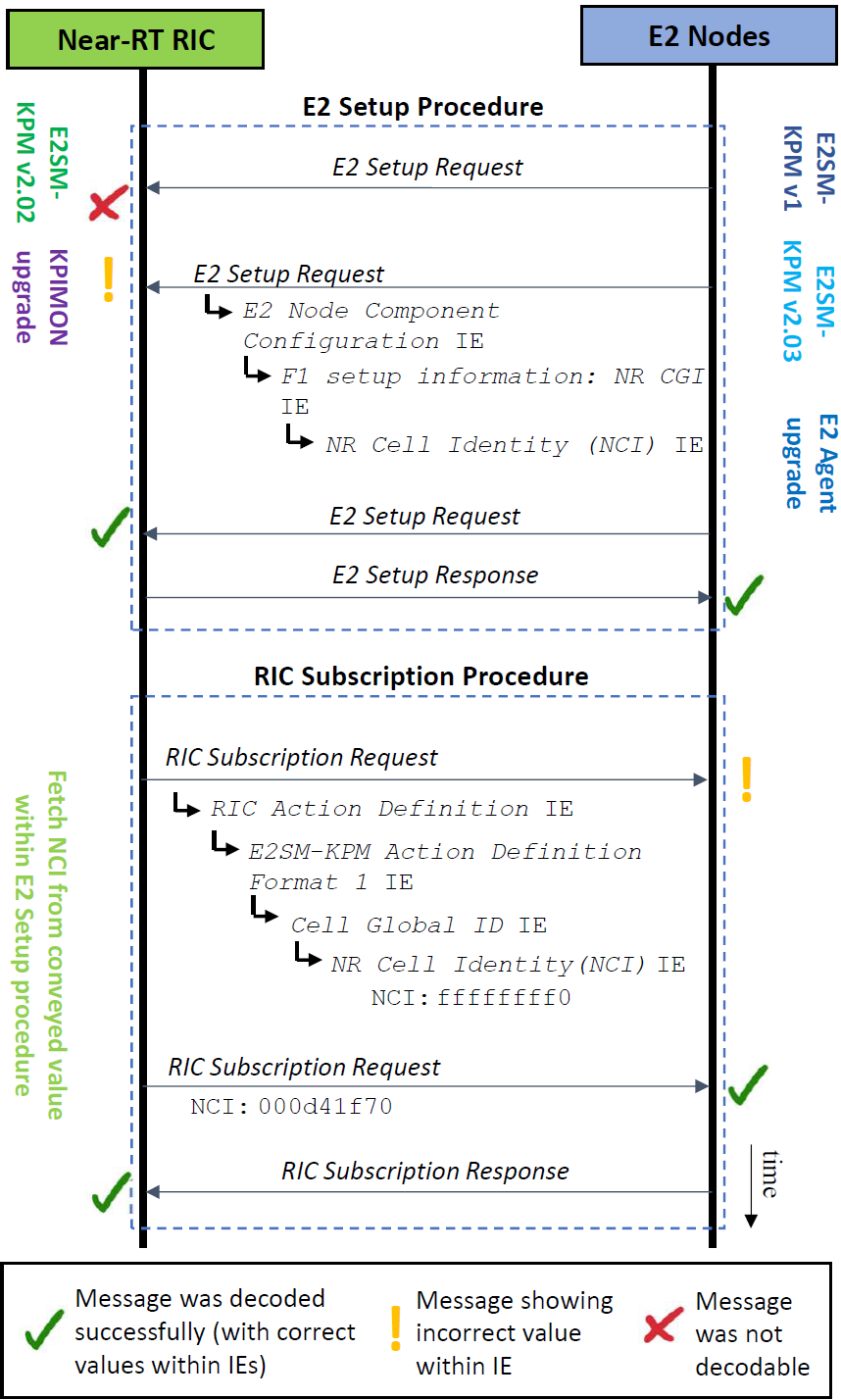}}\caption{Signalling flow between E2 Nodes and Near-RT RIC for E2 Setup and RIC Subscription procedures. Nested IEs need to be validated within IOT procedures to ensure correct interoperability between the RAN nodes.}
\label{ricsignalling}
\end{figure}

After conducting the root-cause analysis within the nested IEs (as illustrated in Fig. \ref{ricsignalling}), it was found that E2 Agent within the E2 Node was hardcoding the F1 setup information \cite{f1info}, including the NR Cell Group ID (CGI) which was conveyed within the \textit{E2 Node Component Configuration} IE of \textit{E2 Setup Request} message from the E2 Nodes. Here, NCI was specified within \textit{NR CGI} IE which was incorrect value conveyed to the Near-RT RIC as part of this message exchange and could finally causing errors in messaging between E2 Nodes and Near-RT RIC. This was resolved by altering the E2 Agent implementation within the E2 Nodes to dynamically process the NR CGI information. 

Although the E2 Setup procedure was successful, the E2 Agent within the E2 Nodes was not correctly decoding the \textit{RIC Subscription Request} coming from the Near-RT RIC. After analysing the messages, it was found that the Near-RT RIC was using incorrect NCI (ffffffff0) within the \textit{RIC Subscription Request} message instead of using the correct NCI (000d41f70). As shown in Fig. \ref{ricsignalling}, this value was specified within \textit{RIC Action Definition} IE, which provides additional information for the nominated Report \cite{E2SMKPMv203}. This observation showed that the KPIMON was not reading the real NCI conveyed within the latest \textit{E2 Setup Request} message, rather it was using hardcoded NCI within the Near-RT RIC software instance. Following the upgrade of software implementation, the RIC Subscription procedure was successfully verified.   

These two examples were not the only cases in which NF implementations for Open RAN components (e.g. E2 Agent) were upgraded as a result of observations from conducting IOT verifications, and similar issues can be found in different domains of interoperability assessment. For example, although the Near-RT RIC was successfully decoding the \textit{E2 Setup Request} message sent from both O-DU and O-CU, the E2 Manager within the Near-RT RIC was responding the \textit{E2 Setup Response} message according to its hardcoded \textit{Global gNB ID} and \textit{gNB-DU ID} values instead of the identities of the connected E2 Nodes. Following an upgrade of E2 Manager implementation to fetch the E2 IDs from the \textit{E2 Setup Request} message, the E2 Setup procedure was successful.

Through these examples, it was shown that the successful verification of one IOT test case and validation of the related procedures does not necessarily translates into successful interoperability between the pairs of components, and it requires further analysis within IE data structures. This is particularly crucial with respect to design implementation of E2 Agent (within E2 Nodes) and E2 Manager (for the Near-RT RIC). 

\section{Concluding Remarks}
Open RAN is now deemed to be a dominant architectural framework for the future mobile radio networks and therefore, the interoperability assessment of Open RAN solutions in commercially neutral testbeds becomes increasingly important. In this paper, we shared insights about the lessons learned from the interoperability assessment of Open Fronthaul and E2 interfaces within Open RAN systems that were integrated in SONIC Labs. We are extending the analysis of the experimental evaluation results, and we plan to publish further results and key findings in the future.


\end{document}